\documentstyle[12pt,epsfig]{article}

\newcommand{\be}{ \begin{eqnarray}}
\newcommand{\ee}{\end{eqnarray}}

\title{ THE PHASES OF QCD\footnote{Summary talk at RHIC Summer
  Studies, Brookhaven, July 1996}}
\author{ E. SHURYAK \\
 Physics Department, State University of New York\\Stony Brook, NY 11794}
\begin{document}
%\noindent{ Physics Department, State University of New York, Stony
%Brook, NY 11794}
\maketitle
\vskip .5cm
{\bf Abstract. }{\small In the recent years we have learned that light quarks play
a crucial role in QCD-like theories, transforming it to many
different phases. We review what is known about them, both from lattice
and
non-lattice approaches. A particularly simple
mechanism of the QCD chiral restoration  phase transition is discussed first:
it suggests that it is a   transition  from randomly placed tunneling
  events (instantons) at low T to strongly localized tunneling-anti-tunneling
pairs at high T. Many features of the transition found on
the lattice can be explained in this simple picture. Very relevant for RHIC, 
this approach predicts a strong non-perturbative interaction
between quarks  $above$ the phase transition. It also predicts that
QGP-like phase sets in at $zero$ temperature, provided few more light
quark flavors are added to QCD. Finally, we 
also discuss possible experimental signatures of the QCD phase transition.
One issue is CERN dilepton data, possibly related with
``dropping'' masses of $\rho, A_1$ mesons. Another is
direct manifestation of a $softeness$ of EOS (smallness of 
pressure/energy density)
in the phase transition region in flow and even the global lifetime of
the system.
}
\section{Introduction}

  My topic is to review our current understanding of critical
  phenomena
in QCD (and its close relatives). But before that, let me comment on 
a somewhat different type of critical phenomena discussed in one (late)
session
of the workshop, the so called ``self-organized criticality''
exemplified by the famous sand piles. Not
being an expert in this field, I still 
dare to suggest that the RHIC summer workshop
itself is a perfect example of this phenomenon. Indeed, only several years
ago many of the participant of RHIC project 
(experimentalists, accelerator people and even most of the theorists)
would not even listen to a
talk with such title as mine. (This is commonly described by
exponentially decaying correlations, with rather short correlation
length.) Now, with common goals and concerns about directions
future
experiments at RHIC will follow, we are all 
 well inside ``one correlation length''. 
This is clearly a phase transition in its own right.
   
   Returning to QCD, let me start with the comment
   that during the last few years we have learned about many
    new phases,  which the gauge theories may have. Of course, the one
    we are going to look for experimentally
 is still the {\it chirally symmetric
Quark-Gluon Plasma} (QGP), in which the charge is $screened$ \cite{Shu_78}
rather than confined. Recent unexpected findings suggest that it
may have completely new features, such as preserve some hadronic modes
as a bound states \cite{SS_95}. 

  However, theory predicts some other phases which has taught us many
  new
lessons. In particular, it seems that
QGP  may exist even at $zero$ temperature in
QCD with about 5-7 light fermions. 
  Even more flavored QCD, with 7-16 light quarks, 
is expected to be in even stranger $conformal$ phase, an extended
relative of a condition  commonly studied only at the second order phase
transition points. Similar phases are proven to exist in
supersymmetric extension of QCD, in which there were recently a
significant advances due to Seiberg and Witten.

  In addition, we have learned  about some $unwanted$ phases, such as
  (i)
the {\it Aoki phase} appearing
in
lattice studies with Wilson fermions
(see Ukawa review) and (ii) the {\it Stephanov phase} appearing in quenched
lattice simulations with the non-zero chemical potential.

As emphasized by T.D.Lee in his opening talk, before thinking about
``small''
(the particles) one should first clearly understand ``large'' (the
phase one actually is in): as we will see
it is a very good advice to these lattice
calculations indeed.

  The bottom line of this talk
is the crucial (but still not quite well understood) role of light fermions,
generating all these phases of QCD. In particular, 
  compare the phase
transition found in 
``quenched'' calculations ($N_f=0$) and those with $N_f=2-3$
dynamical quarks. The former case has a ``deconfinement''
phase transition at  high $T_c\approx 260 MeV$, while 
the latter show a ``chiral restoration'' transition
 at much lower $T_c\approx 150
MeV$ \footnote{ By changing the quark masses from
light to heavy continuously, it was  shown that these two transitions
are
indeed different phenomena, 
separated by a large gap in which there is no transition at all.}
 Unlike at normal(T=0) conditions, at $T\approx T_c$ physical
 quantities are
very sensitive to such little details as
 the mass value of the strange quark. Therefore we are
 still
not quite sure about the $order$ of the transition: the latest lattice
results \cite{Kan_96} incline again toward the 1-st order in the real
world.

 But do we understand
 why the transitions happen? Is there a simple picture which can
 explain its
microscopic mechanism? Can we build a working model, reliable enough
to  provide some guidance in  delicate questions relevant to 
 experimental observables?

   Qualitative explanations of why  the QCD phase transition takes place  
are often done in an over-simplified way,  emphasizing the ``overlapping''
hadrons in a  bag-model-type picture. But such pictures give all
numbers
 \footnote{For example, the MIT bag model literally predicts  QGP formation in
   heavy ion collisions at
 unrealistically small  (BEVALAC/GSI) energies.} and physics wrong.
Pure gluodynamics
is an especially good example:
 at  $T_c\approx 260 MeV$ the density of glueballs is  negligible
  since even the lightest one has a mass of about 1.7 GeV! Looking
  at the low-T side of this transition it is impossible to tell why
it happens. The same is qualitatively true 
for the chiral restoration: on the hadronic side 
 the matter is still relatively  dilute. But the reason
why the phase transition happens is very simple: just two very different
phases happen to
 have {\it the same free energy}. The lesson: one cannot understand a QCD
 phase transition without a $quantitative$ model for $ both$ phases. 

   The energetics of the  ``deconfinement'' and ``chiral restoration''
   transitions is  entirely different, suggesting
different physics. The former has huge
   latent heat,  few $GeV/fm^3$, so that probably all of the
 non-perturbative
vacuum energy density (proportional to the ``gluon condensate'')
is ``melted'' in it. This in not the case for ``chiral
restoration'' in QCD:  large portion  of the gluon condensate
should actually survive it \cite{KB_93}. 
What is this remaining ``hard glue'' or ``epoxy'', as  Gerry Brown
called it? Why, unlike ``soft glue" does it  not
 produce a quark condensate? Can it affect quark interactions and
hadronic masses? 
   
   The major
``non-lattice'' approaches to the problem include: (i) models based on 
particular effective Lagrangians, like the sigma model or chiral effective
Lagrangians 
models \cite{Pisarski_95,HK_94};
 (ii)  QCD sum rules at finite temperature/density \cite{HSK_96}; (iii)
the interacting instanton liquid model (IILM) \cite{SS_95,SVel_96}.
%  IILM is based on very specific dynamics, but
%it is superior to others because of its
%more microscopic base and also because its results reproduce
%surprisingly well the (T=0) phenomenology and
%the lattice results. But let me make brif remarks first on its rivals.

  {\it Effective Lagrangians} are very powerful tools at $low$
   temperatures:
with parameters fixed by data one can
indeed accurately account for effects due to non-zero occupation
factors. However, they are clearly ``one-sided'', unable to deal with
  QGP.
In principle, {\it   QCD sum rules} are  $not$ limited to a
description of only hadronic phase:  the structural changes 
can be adequately  described by VEV of different
operators, or ``condensates''. 
However, since in practice those are unknown,  people  use
simplifications. The most popular one is
``vacuum dominance'',  
reducing average values of any quark operators to  powers of  $<\bar q q>$,
  therefore missing  non-perturbative
effects above $T_c$.

    Among theory issues discussed intensely during the last couple of
    years
\cite{Shu_94,Cohen_u1,KKM_96,Wang_96} is the fate of the U(1) chiral
symmetry. In cannot be exactly restored, as suggested in
\cite{Cohen_u1}, but its violation certainly is dramatically reduced
at $T\approx T_c$. In practice it means
that $\eta_{non-strange}$ (a combination of $\eta,\eta'$) and
isovector scalar (we call $\delta$) may
  be nearly as light as a pion.

 One very important issue (which should have been discussed more)
is what happens with the QCD phase transition for larger
number of quark flavors $N_f$.  Several phase transition lines
are expected there, the lowest probably being  the chiral
symmetry restoration at T=0.

\section{A mechanism for the chiral phase transition}

  The main point of this talk is that due to recent developments it
  becomes increasingly clear what is the
 {\it microscopic mechanism} underlying this phase transition. It is
$rearrangement$ of instantons, from relatively random liquid at low T to 
a gas  of  instanton anti-instanton
``molecules''\footnote{Note a similarity to  Kosterlitz-Thouless
  transition in O(2) spin model in 2 dimensions: there are paired
  topological objects (vortices) in one phase and random liquid in
  another.}. These $\bar I I$ molecules are the
 ``epoxy'' mentioned above: at $T>T_c$ they indeed
do not create the $quark$ condensate but contribute to the $gluon$ one.
Furthermore, they generate new type of
 the inter-quark interaction, and may even create
hadronic states, even above $T_c$!

   Several events  lead to these developments: (i)
Demonstration that ``instanton vacuum'' explains the QCD correlation
functions and hadronic spectroscopy at T=0
\cite{SSV_94,SS_95};
 (ii) Lattice confirmation of instanton liquid parameters
and of the dominant role in general\cite{CGHN_94,MS_95b} ;
(iii) As shown in \cite{SVel_94}
 instantons cannot be screened\footnote{Previously considered
scenario based on the ``instanton
   suppression''  does not in fact work until l
   rather high T, well in the QGP domain, and thus it cannot be the reason
   for the phase transition.}
  at $T<T_c$; (iv) Confirmation of
 this statement on the lattice \cite{CS_95}; (v) Discovery of
 polarized  $\bar I I$ molecules at $T\approx T_c$ \cite{IS_94};
 (vi) Numerical simulations \cite{SS_95}
and analytic studies \cite{SVel_96} of the phase transition
in the interacting instanton ensemble.

\begin{figure}[t]
\begin{center}
\includegraphics[width=6.cm]{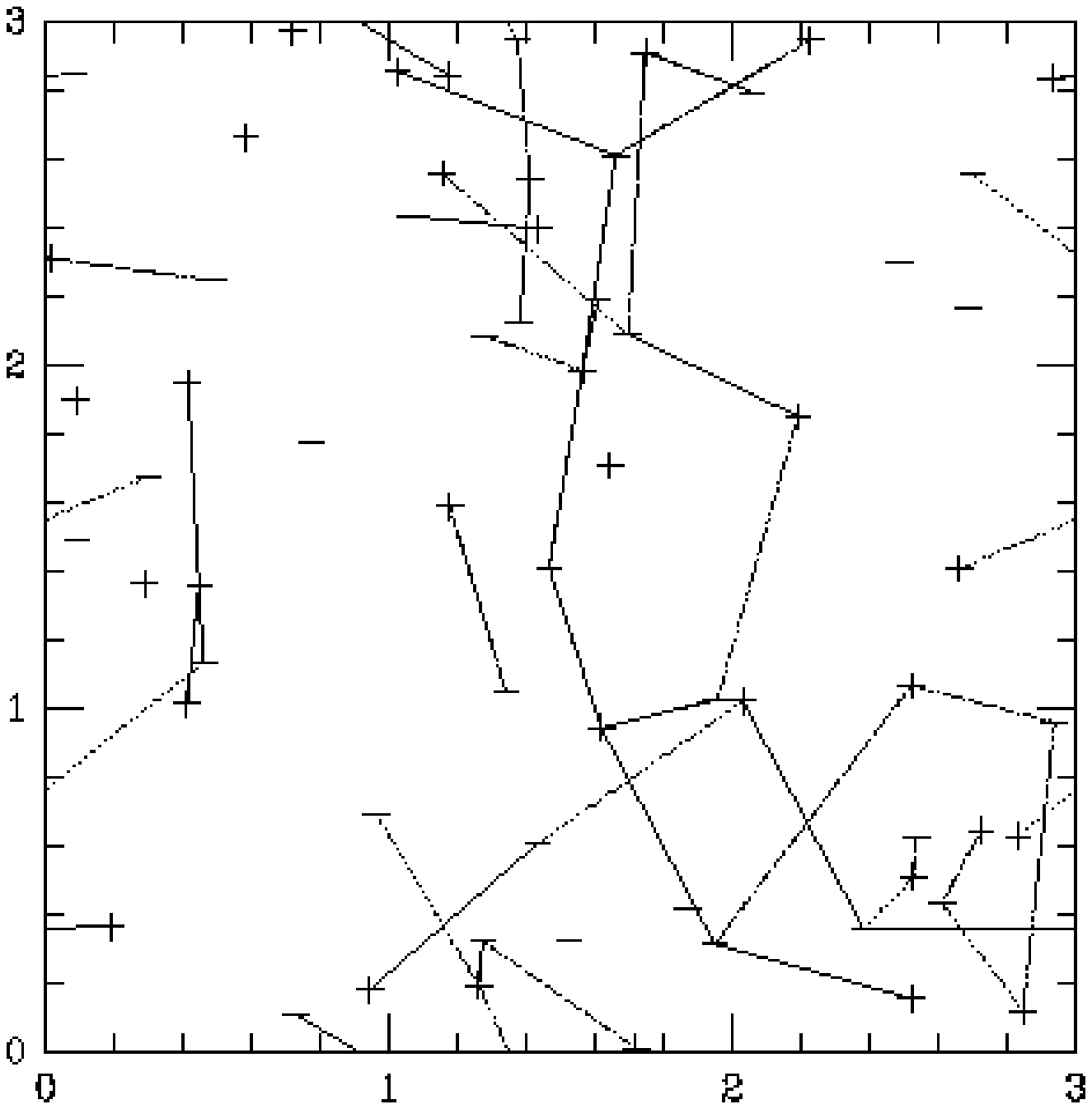}
\hspace{1cm}
\includegraphics[width=6.cm]{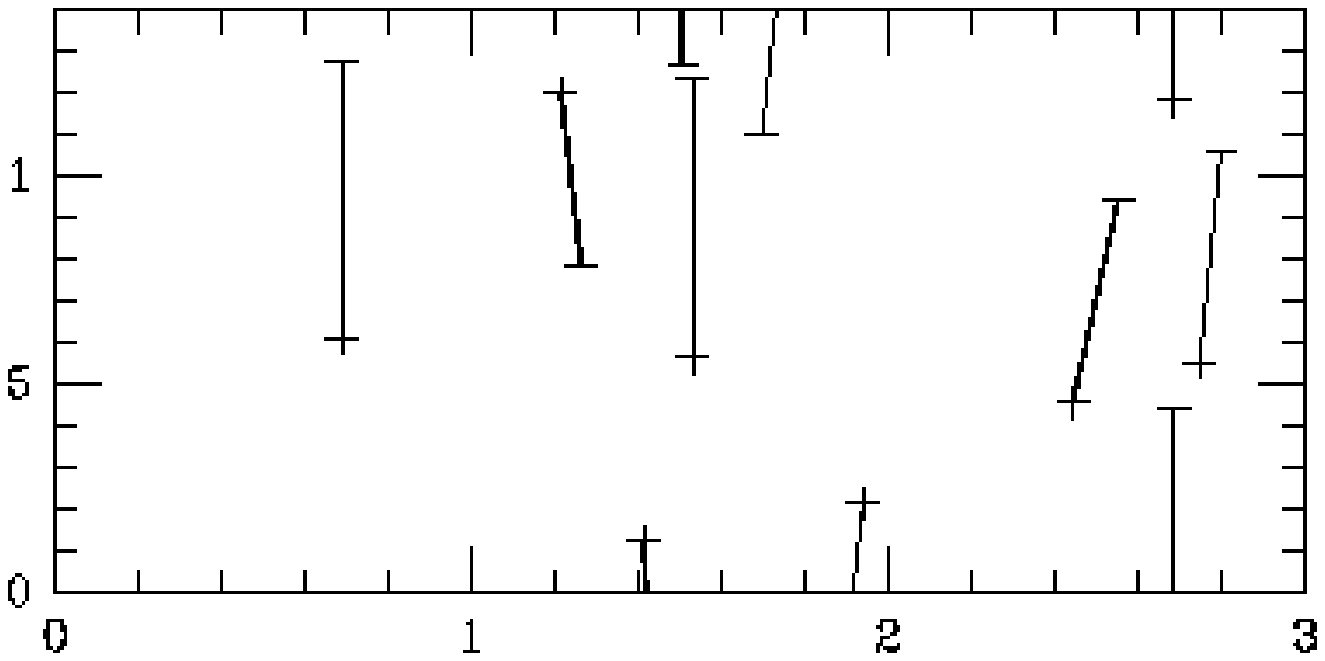}
\end{center}
\caption{\label{molecb}
Typical instanton configurations for temperatures
T= 76 and 158 MeV.
}\end{figure}

   It is very easy to explain  what happens at $T\approx T_c$ in this
   approach. Recall that finite temperature is described in Euclidean
   space-time by
periodic boundary conditions, with the Matsubara period $1/T$. 
So a rising T means a decreasing box, and when it fits to
the size of one  $\bar I I$ molecule, one gets a ``geometric'' transition.
In   Fig.
\ref{molecb} from\cite{SS_95} one can see it clearly. 
(The plots show projections of a four dimensional
 box into the $z$ axis-imaginary time plane.
Instantons and anti-instanton positions are indicated by $+$ and $-$ symbols.
The lines  correspond to strongest fermionic ``bonds''.)
 Notice that molecules
are strongly
``polarized'' in the time direction 
 and that they are separated by half Matsubara box in time
$\Delta\tau=1/(2T)$.\footnote{  Thus one can even find the approximate
 phase transition 
temperature. The molecule  fits onto the torus if $4\rho
\simeq 1/T_c$, and with known instanton size $\rho\simeq 0.35$ fm 
one gets $T\simeq 150$ MeV, close enough to the observed one. }

In a series of recent numerical simulations \cite{SS_95} 
it was found that like QCD,
 the instanton model has $second$ order transition for
$N_f=2$ massless flavors, but a {\it weak first order} one for QCD
with physical masses. 
 Furthermore, the thermodynamic parameters, the
spectra of the Dirac operator, the T-dependence
 of the quark condensate and various susceptibilities, 
the screening masses 
  are all
consistent with available lattice data.

%\twocolumn

\begin{figure}[h]
\begin{center}
%\includegraphics[width=3.cm]{pion.ps}
%\hspace{1cm}
\includegraphics[width=15.cm]{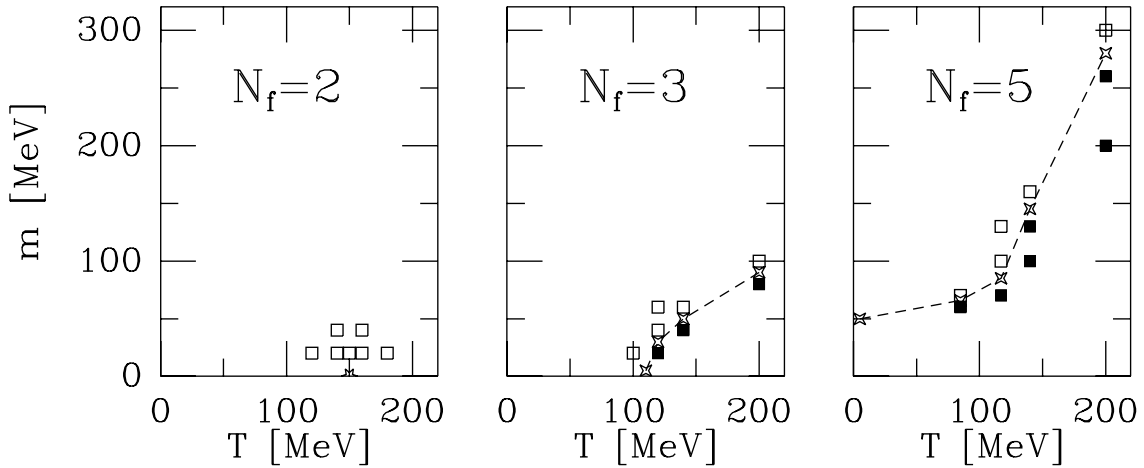}
\end{center}
\caption{\label{pion}
Phase diagram (temperature-quark mass plane) of the instanton liquid for
different numbers of quark flavors, $N_f$=2,3 and 5. 
The open squares indicate
 non-zero
chiral condensate, while  solid one indicate that it
is zero. The dashed lines 
show the approximate location of the discontinuity line.
}
\end{figure}

\section{Phases of QCD with more quark flavors}

\begin{figure}[t]
\begin{center}
\includegraphics[width=15.cm]{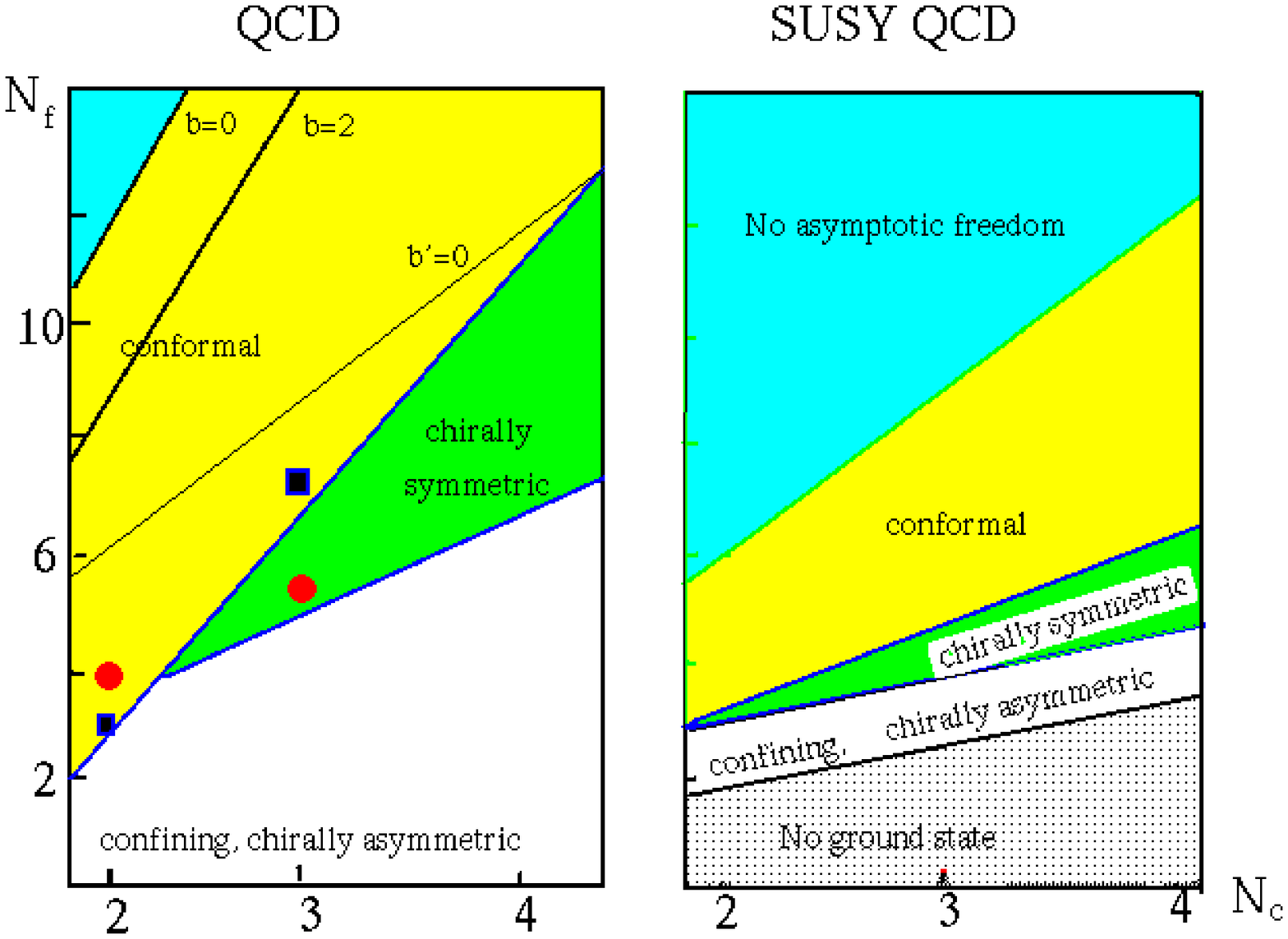}
\end{center}
\caption{\label{fig_map} 
Schematic phase diagram of zero T QCD  (a) and supersymmetric QCD (b)
as a function of the number of colors $N_c$ and the number 
of flavors $N_f$.} Squares show where lattice calculations have found
the
infrared fixed point, dots are where the instanton ensemble is
purely ``molecular'', with the unbroken chiral symmetry.

\end{figure}

  In this section we  discuss further the role of quarks in QCD,
  adding
more flavors to it. 
  If we add too much of them, namely
 $N_f>33/2$ (here and below we imply 3 colors), the asymptotic freedom is
  lost and we get  $uninteresting$ field theory with a charge growing
  at
small distances, basically
a theory  as bad as QED! So our ``most flavored'' QCD (with 16
  flavors,
and work out way down from it (see fig \ref{fig_map}). 
The phase we are in there is actually rather simple one, known as 
 the Banks-Zaks conformal domain \cite{BZ_82}. It has 
 the infrared fixed point at small coupling $g^2_*/16\pi^2=-b/b' <<1$
($b,b'$ are the one and two-loops coefficients of beta function). It happens
 in the perturbative domain, so the
charge is small both at small and large distances. There are no
particles in this phase, and all correlators decay as powers of the
distance. In this phase
the non-perturbative phenomena like instantons are exponentially
suppressed, $exp(-const/g^2_*)$. However, as one decrease the fermion
number,
the fixed point $g^2_*$ moves to larger values and eventually
disappears. 
   Lattice simulations of multi-flavor QCD were recently reported 
in \cite{IKK_96}. These authors studied QCD with up to 240 flavors. 
 Studying the sign of the beta function in the weak and 
strong coupling domains, they confirmed the existence of an infrared 
fixed point
as low as  at $N_f=7$. 

  It is not known which phase we find next, at $N_f=5-6$,
most probably it is the so called Coulomb phase (basically QGP).
 
The results of the interacting instanton model are summarized by
 Fig.\ref{pion}(b-d) show how
one singular point at $N_f=2$ develops into the discontinuity line
for $N_f=3$. The value of $T_c$ goes down with increasing $N_f$ and at
$N_f=5$\footnote{Note
that $N_f=4$ case is missing. It is because I have found the
condensate to be small and comparable to finite-size effects. In order
to separate those one should do calculations in different boxes, which
is
time consuming.} one finds that
the
 chiral symmetry is restored even at T=0, provided quarks are light enough.

  New lattice results have been presented at this meeting by R.Mawhinney
for
QCD
with  $N_f=4$. Details can be found in his (and N.Christ's) talk, and I
should only say that when they have extrapolated the measured masses
to  quark masses $m\rightarrow 0$,
 they have found a dramatic significant drop in chiral symmetry
 breaking effects, such as $\pi-\sigma,\rho-a_1,N-N^*(1/2^-)$
 splittings, very much unlike the
$N_f=0-3$ studied before. It suggests that chiral restoration 
is nearby, 
very similarly to what was found in the instanton calculations. 

 Finally, we return to Figure \ref{fig_map}, and explain its rhs, showing
 similar phase diagram for the
 N=1  SUSY QCD based on \cite{Sei_94}.
Not going into details, let me only mention that the vacuum is now
definitely dominated by instanton-antiinstanton molecules, and their
contribution can be calculated without problems (in QCD subtraction of
perturbation theory is a great one). Thre are two phases which are
impossible in QCD: a case without the ground state (molecule force the
system toward infinite Higgs VEV) and also a funny situation with 
chiral symmetry unbroken but confinement (hadrons exist but are degenerate
in parity).
 It is amusing that
chiral symmetry is restored
at $N_f=N_c+1$, similar to what was  found
in QCD in
the instanton model. Also note, that in this case the existence of the
Coulomb
phase is a proven fact.

\section{ Inter-quark interaction and hadrons near and above $T_c$} 
 
   It is well known that the nucleon effective mass in nuclear matter is
reduced, and
there are also  indications that $m_\rho$ 
does the same \cite{rhoinnuclei}. Among
approaches
 predicted ``dropping masses'' at high T  the simplest
(and the most radical) one is  Brown-Rho scaling. According to it
all hadronic masses get their scale from $<\bar q q>$, and therefore vanish  
at $T\rightarrow T_c$. This idea is supported by the QCD sum rules, provided  the
``vacuum dominance'' approximation  is used. However it is not so
in a vacuum made of the instanton molecules, because they generate  
non-zero average values for some fermionic operators, even when
$T>T_c$
and  $<\bar q
q>=0$. Those operators can be treated as 
new non-perturbative
inter-quark interaction, which can be 
 described by the following
 (Fierz 
symmetric, color-singlet only) Lagrangian  \cite{SSV_95}
\be
\label{lmol}
 {\cal L}_{mol\,sym}&=& G
     \left\{ \frac{2}{N_c^2}\left[
     (\bar\psi\tau^a\psi)^2-(\bar\psi\tau^a\gamma_5\psi)^2
      \right]\right. \nonumber \\
     & & - \;\,\frac{1}{2N_c^2}\left. \left[
     (\bar\psi\tau^a\gamma_\mu\psi)^2+(\bar\psi\tau^a\gamma_\mu\gamma_5
     \psi)^2 \right] + \frac{2}{N_c^2}
     (\bar\psi\gamma_\mu\gamma_5\psi)^2 \right\} + {\cal L}_8,
\ee
which has Numbu-Jona-Lasinio-type form and  the coupling constant
is proportional to density of molecules
G = $\int n(\rho_1,\rho_2) d\rho_1 d\rho_2 \,
	\frac{1}{8T_{IA}^2} (2\pi\rho_1)^2 (2\pi\rho_2)^2.$
Here, $n(\rho_1,\rho_2)$ is the tunneling probability for the 
IA pair and $T_{IA}$ is the corresponding overlap matrix element,
$\tau^a$ is a four-vector with components $(\vec\tau,1)$. The 
effective Lagrangian (\ref{lmol}) was determined by averaging 
over all possible molecule orientations. Near the phase transition,
molecules are polarized and all vector interactions are 
modified according to $(\bar\psi\gamma_\mu\Gamma\psi)^2\to
4(\bar\psi\gamma_0\Gamma\psi)^2$.

\begin{figure}[t]
\begin{center}
\includegraphics[width=6.5cm]{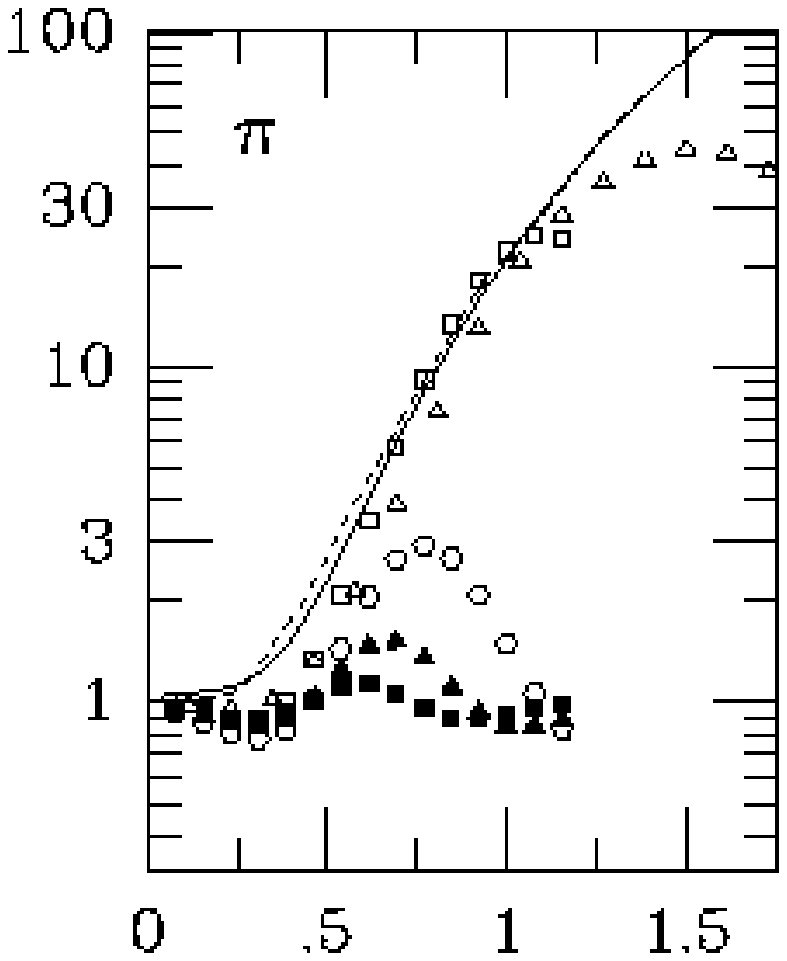}
\hspace{1cm}
\includegraphics[width=6.5cm]{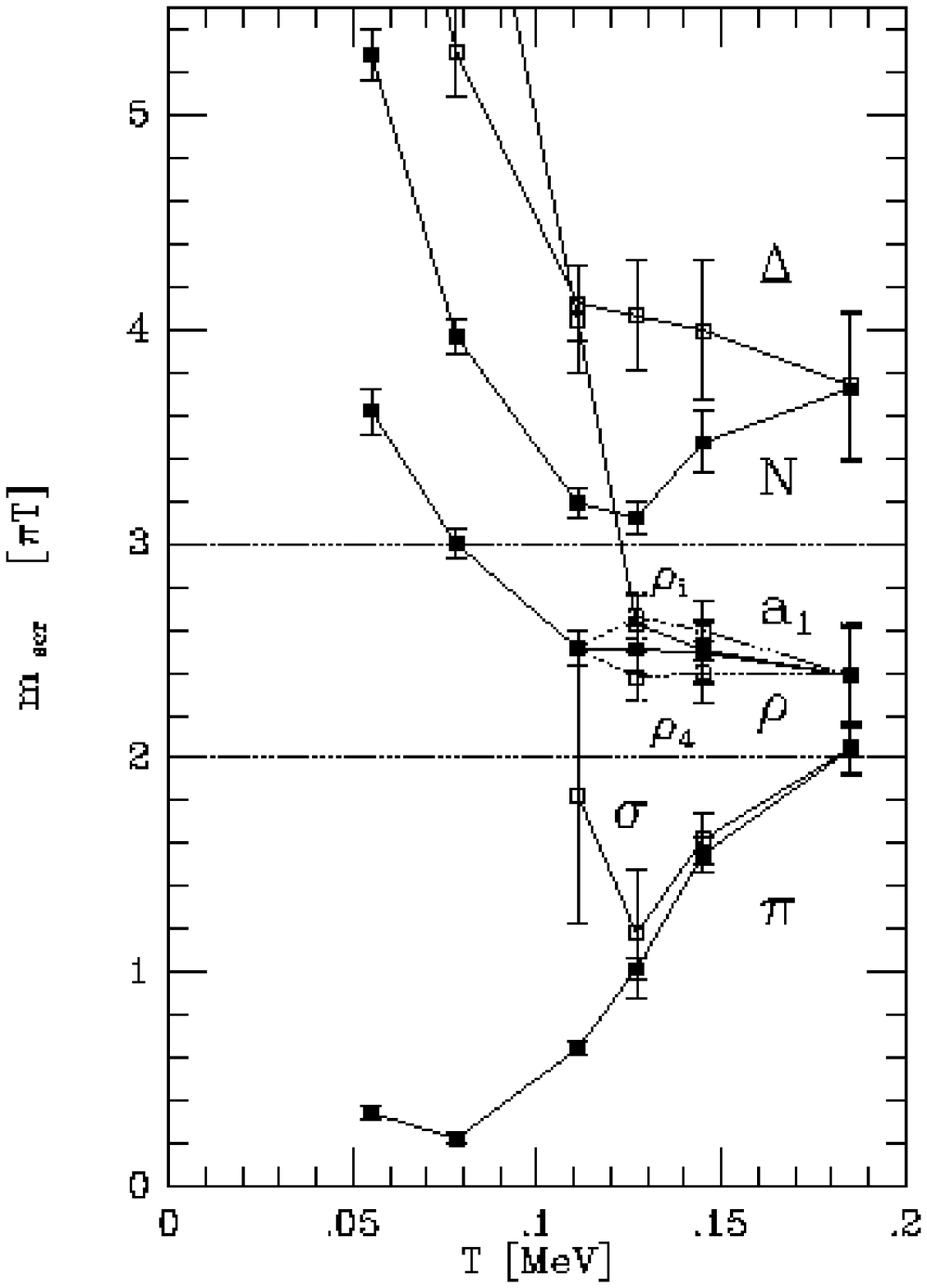}
\end{center}
\caption{\label{QCDcond}
(a) Temporal correlation function for the pion (divided by the one
corresponding to free massless quarks) versus distance (in fm). Solid
triangles and squares are for $T=1.,1.13 T_c$, open triangles, squares
and hexagons for $T .43, 0.6, 0.86 T_c$, respectively.
 (b) T-dependence of
the screening masses (given in units if $\pi T$)
  calculated from the IILM. Note 
that at chiral restoration the partners $(\sigma,\pi),(rho,a_1)$
 become identical.
}\end{figure}

  Numerical simulations\cite{SS_95}  and analytic studies
  \cite{SV_96}\footnote{Those are directly analogous to the BCS theory of
  superconductivity,  but with the instanton-induced interactions like 
(1).}
 have been used to calculate
both  $spatial$ and $temporal$  
correlation function. The former exponentially decay with
the so called ``screening masses'': their 
T-dependence for a number of hadronic channels is shown
in Fig.(\ref{QCDcond})(b) 
\cite{SS_95}  show overall agreement with lattice
ones. Especially important is strong attraction in scalar-pseudoscalar
channels, shifting these masses down from their high-T asymptotic,
$M/\pi T=2$.
This attraction is also clearly seen in
  temporal correlation functions as well. For example, the pion correlators
at several T are
shown in Fig.(\ref{pion})(a). They all have upward enhancement, which is
absent
in other channels (such as $\rho$). Fit of the correlator suggests that
 pion survives the phase transition  and exist at $T>T_c$ as a
 (non-Goldstone)
massive bound state.  Most other hadrons 
``melt" into constituent quarks. Those have no effective mass but 
effective energy (or ``chiral mass") above $T_c$. Unfortunately,
accuracy
of the calculations done so far does not allow to get information
about
the fate of vector and axial mesons, the major players in the next section.

 Unfortunately, the temporal correlators are very difficult to get
 on the lattice: there are  only few points in
 time direction. The best data we can compare
 with  \cite{BGK_94} definitely show attraction in the pion channel 
 at $T>T_c$, 
which is absent in vector ones.

\section{The $U(1)_A$ restoration?}

    The fate of the $U(1)_A$ anomaly at finite temperature has
received a lot of attention lately. A number of authors have
emphasized that a (partial) $U(1)_A$ restoration may be the case, with
rather dramatic observable consequences \cite{Shu_94,KKM_96,HW_96}. 
Indeed, recall that at T=0 the $\eta'-\pi$ mass difference is 
larger than all other meson mass splittings, and any tendency for it
to shrink would therefore
strongly affect the whole spectroscopy\footnote{An interesting situation arises for $N_f\geq 3$ 
massless flavors. In this case, the anomaly does not affect 
the $\eta'$ correlation function above $T_c$ \cite{SS_96b,LH_96,EHS_96}. 
If chiral symmetry is unbroken, extra zero modes cannot be absorbed
by the condensate, and the 't Hooft vertex only contributes 
to $2N_f$-point correlators. For $N_f=3$ this means that the 
$\eta'$ and $\pi$ are degenerate above $T_c$, but the singlet
and non-singlet $\Lambda$ are split.}.

 It is well known that this splitting and the  $U(1)_A$ anomaly are
related with the  topological charge, or well 
 isolated instantons.  It was argued in \cite{Shu_94} that those 
should be very rare
in the ensemble at $T>T_c$, because all instantons are ``paired''
into molecules. 
The density of isolated instantons 
 above $T_c$ is small $O(m^{N_f})$, but due to
 zero modes the quark propagators contribute  
a factor $1/m$, and some results are finite in the chiral limit\footnote{
Another way to explain it is to say, that for m=0 the individual
instantons are absent in the vacuum, but can be created by the
operators themselves.}. So, a significant drop in the strength of
the $U(1)_A$ anomaly around $T_c$ is expected, although 
to non-zero value.
  It was further suggested in \cite{Shu_94} that instead of dealing
  with $\eta'$ (which involve a complicated flavor mixing pattern), it
  is more convenient to look at a difference between $\pi$ and its 
 $U(1)_A$ partner isovector scalar $\delta$. 

   Such measurements have been done, both on the lattice and in the
   instanton model. $All$ of them observe that the difference
of   $\pi-\delta$ susceptibilities indeed drops dramatically at $T_c$,
 indicating a move toward the $U(1)_A$ restoration.

 However, the
 calculations
disagree about what exactly is its value at $T>T_c$. The
problem is the $m\rightarrow 0$ limit, which is very difficult to do
in practice. Then the conclusion is sensitive to extrapolations:
  MILC  collaboration \cite{MILC_96} has fitted it
quadratically
 $\chi_\pi-\chi_\delta=C1+m^2 C2$ and gets a non-zero $C1$, but
 extensive results from the
 Columbia group (see their data in N.Christ's talk) have shown that the
 linear
fit is much better, and it leads to $\chi_\pi-\chi_\delta \rightarrow 0$.

  In the instanton calculations \cite{SS_95} the value is clearly
  non-zero
(although it is not  accurately determined yet). So, 
it seems that here we (for the first time) see a serious
qualitative disagreement with lattice simulations at $T>T_c$.
This can be seen explicitly from the form of the Dirac spectrum. In
the one we
have calculated there are still quasi-zero modes due to isolated
instantons,
while in Columbia ensemble those seem to be completely absent. 
Why this happens remains unclear.

The problem of $\eta\eta'$ mixing at finite T was studied in more detail
in \cite{Sch_96}. The most important conclusions is that there
is a tendency towards ideal mixing, a separation of strange and
non-strange component. However,
contrary to others, it is shown that for $T>T_c$ the anomaly
only can operate in the   non-strange sector, with $\bar u u$
transition into $\bar d d$ being proportional to $m_s$. It is claimed
that non-strange component may be rather heavy.

In summary, a clear tendency toward diminishing of the role of U(1)
anomaly
at $T\approx T_c$ is observed, but quantitative results for  $T> T_c$  
are still missing. If (as Columbia data suggests) it is vanishingly
small, we would have 8 rather then 4 massless modes (for $N_f=2,m=0$),
 and understand why it moved away from O(4) indices to the first-order ones. 
However, the instanton calculations show that it is not so small.

\section{The phase transition at finite baryon density: the old and
  the new puzzles }
 
   First attempts to introduce the non-zero chemical potential
   associated with baryon number have been done in $quenched$
   approximation a decade ago \cite{Bar_86}. It was  found
that a very
   strange thing (referred as the ``old puzzle'') happens. At small
   $\mu$ as expected,
   nothing depends on it, till some threshold $\mu_c$ is reached. But
instead of finding a threshold to be around the constituent quark mass
 $\mu_c\approx m_N/3 $ \footnote{By
  tradition, lattice people just put $\mu$ in the Dirac operator without
  extra coefficients,
  which means that in their units quarks have baryonic charge 1, not
  1/3.},
they have found it at  $\mu_c\approx m_\pi/2$. Those are not too
different if quark bare mass is large, but if it goes to zero the
latter value vanishes. That means that 
in the chiral limit some states with non-zero
baryonic number exist (and can be excited) at arbitrarily small $\mu$.
There are no such states in the real world, so why lattice data have
found them?

  For ten years it was not clear why, but the phenomenon was not a
  numeric artifact, and it has been since then seen in many other
  calculations.
   The resolution of the puzzle was recently made by Stephanov
   \cite{Ste_96}.
He has pointed out, that it is ``quenching'' the QCD partition function  
which one should blame. Interestingly enough, it turns out that
elimination of the quarks can be done in two very different ways.
In both one writes in the bosonic
partition function the factor $[det(i\hat D +i\mu \gamma_0 +im]^{N_f}$
coming from the integration over fermions, and then takes
$N_f\rightarrow 0$. This factor is real at $\mu=0$ and complex
otherwise, so it can be written as a modulus and a phase.
Stephanov has shown that the usual quenching corresponds to the limit,
in which one keeps modulus and ignores the phase\footnote{One can do
  this also without the limit, at finite $N_f$: such approach was
 suggested and studied
separately by Gocksch \cite{Goc_88}.}. 

  Putting the absolute value of the determinant means that anti-quarks
  have a conjugate operator, with the opposite sign of the
chemical potential $\mu\rightarrow -\mu$: so instead of having the
baryonic charge -1 they also have it equal to 1, like quarks. Then 
 mesons $\bar q \Gamma q$ may have the baryon charge 2, and the
massless excitations excited at small $\mu$ are nothing else but such
pions. Those exist in the Stephanov
(unwanted) phase, in which the baryon number is spontaneously broken by
the quark condensate\footnote{In fact Stephanov has found 
  boundaries of this phase and many interesting details about
  distributions of Dirac eigenvalues at non-zero $\mu$, both in the wrong
(no phase) and right (with the phase) ensembles, using
  a particular model of chiral restoration
 based on random matrix approach. Unfortunately, 
 we have no time to discuss it here.}, which also obtains the baryon charge.

   Another talk we had on the lattice data at finite baryon density
have been given by M.Lombardo. This time it is based on rather
complicated algorithm and strong coupling limit, but it is unquenched
and  includes the determinant as is, with its phase.
 
A good news is the excitation to non-zero baryonic density happens
around the expected threshold value. But
the bad news is  the excitation starts about 30\% lower
it and is more smooth compared to what one expects on the basis of the
temperature value involved. So, we still see some unusual baryonic
states being excited, like some nuclei or nuclear matter
 with unexpectedly large
binding energy. I do not know any explanation of this puzzle.

\section{ ``Dropping masses'' in the experiment?} 

  Already at QM93,  CERN dilepton experiments have provided first
  preliminary indications for a signal, significantly exceeding  
  the theoretical expectations (e.g.\cite{SX_where}). Later CERES 
  has found dramatic excess of dileptons at
  $M_{e+e-}<m_\rho$. 
  In spite of multiple attempts by theorists to explain it
  by ``conventional sources'', it was found to be impossible
\footnote{In fact, I have never before  seen
  that rather involved calculations by several groups agree so well,
  as far as the shape of spectra of $M_{e+e-}$  is concerned. }.
  A list of ``unconventional'' explanations (more or less in
 a chronological order)
include: (i) dropping $m_\rho$
   \cite{Hof_94,LKB_95}; (ii) pion occupation numbers at low momenta are very high \cite{KKP_93}; 
(iii) a very long-lived fireball \cite{SX_where}; 
(iv) dropping  $m_{\eta'}$ \cite{KKM_96,Wang_96}; (v) a  modified pion
   dispersion curve \cite{Song_96}; (vi) dropping $m_{A_1}$ \cite{HS_96}.

  I will only discuss the $\rho-A_1$ story, with  brief
remarks about others. (ii) The observed low-$p_t$ excess of pions (over thermal
spectra) can indeed lead to a dramatic increase in low-mass dilepton
yield. However, it is
 most probably due to resonance decays or effects of
collective potentials. Both are
 late-stage phenomena, which hardly affect the early-stage dilepton 
production. 
(iii) Even if the long-lived fireball would appear at  200
GeV/A (which is hardly possible)
 it was found to produce 
 about the same $M_{e+e-}$  spectrum  as the
usual space-time scenarios, so it does not work. (iv) Above we have indeed 
suggested that $m_{\eta'}(T)$ should drop more
than any other mass: however   in order to account for
 CERES data
one should increase  yield of ``escaping  
 $\eta'$'' by too huge a factor. (In
   connection to U(1) restoration issue, however, it would
 be extremely interesting to measure the  $\eta'$ yield, though.)

  The ``dropping $m_\rho$'' idea is very well known, and   
it seems to be about the only one which can explain the dilepton data. 
It was studied in details by Li,Ko and Brown \cite{LKB_95}
in the cascade model based on Walecka-type model with attraction
mediated by a scalar field.
 With much
simpler hydro-based 
approach  we have also verified, that one can get a very good
description of CERES data
by making the rho mass T-dependent,
$without$ any changes in
 standard thermal rate formulae or in  space-time evolution.

%\begin{figure}[h]
%\begin{center}
%\includegraphics[width=3.cm]{rhoamass2.eps}
%\hspace{.2cm}
%\includegraphics[width=6.cm]{cases.eps}
%\hspace{.2cm}
%\includegraphics[width=6.cm]{cased.eps}
%\end{center}
%\caption{\label{Mspectrum}
%(a) Possible scenarios of chiral restoration in $m_\rho - m_{A_1}$
%plane, (b)  the dilepton spectra corresponding to them (CERES data are
%shown with background from hadronic decays subtracted; (c) the most 
%favorable case D is shown in details, with contribution from different
%stages. The dominant one is clearly the long-dashed one, or hadronic
%part of the mixed phase. 
%}\end{figure}

\begin{figure}[!h]
%%\begin{center}
\begin{minipage}[t]{0.333\linewidth}
\includegraphics[width=6.cm]{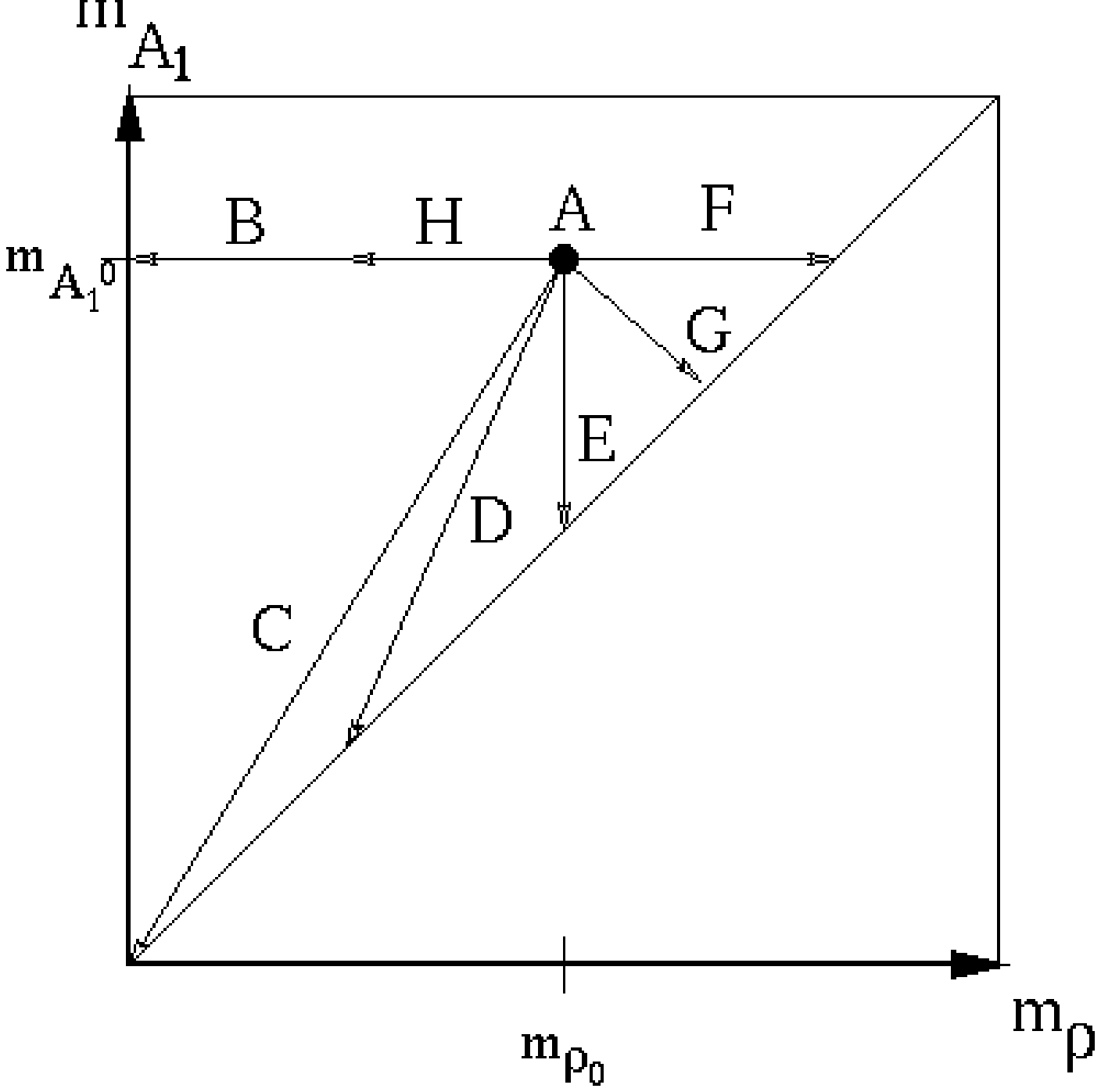}
\end{minipage}
\hspace{1cm}
\begin{minipage}[t]{0.333\linewidth}
\includegraphics[width=6.8cm]{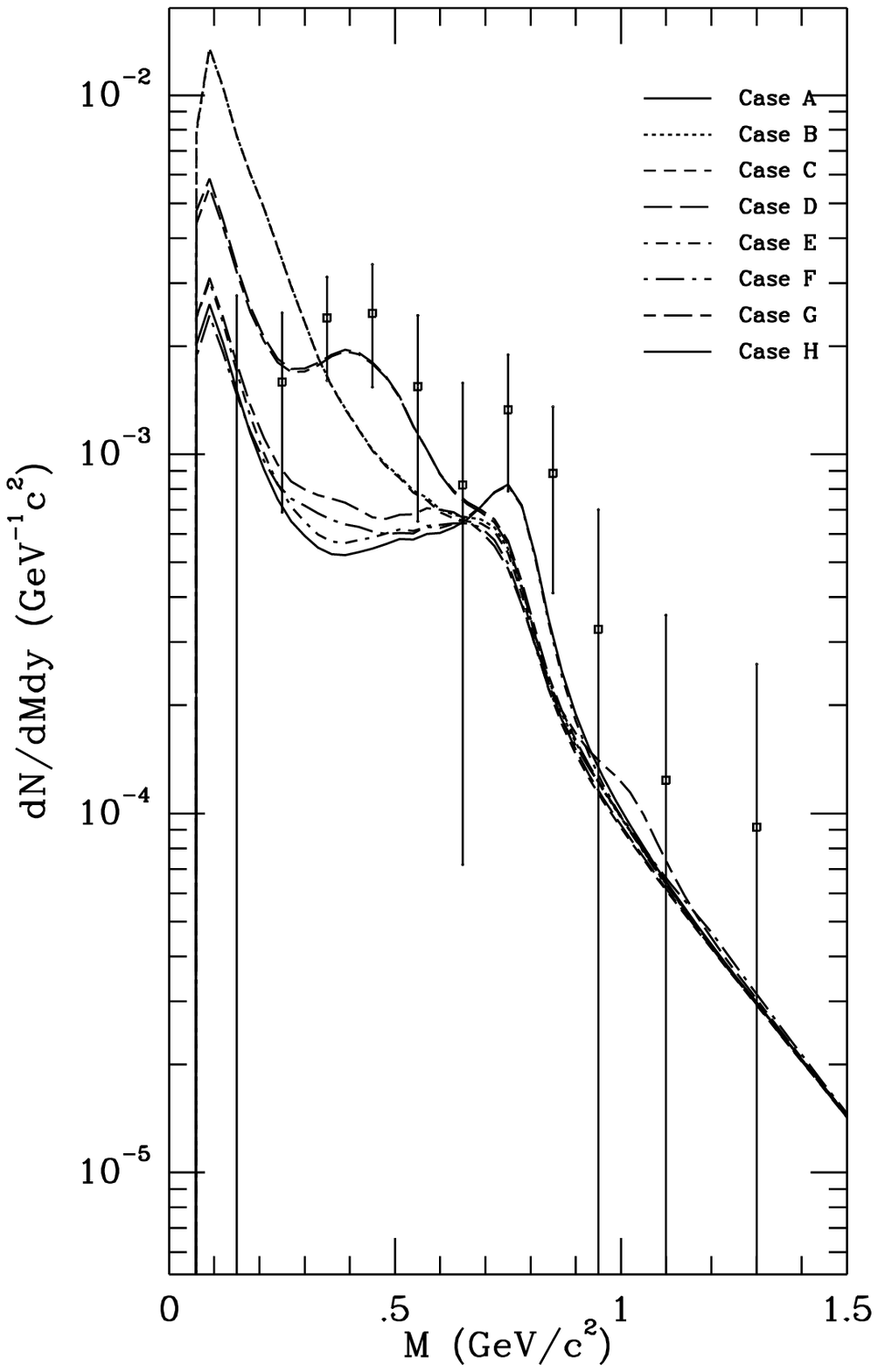}
\end{minipage}
\hspace{-11cm}
\begin{minipage}[t]{0.333\linewidth}
\includegraphics[width=5.7cm]{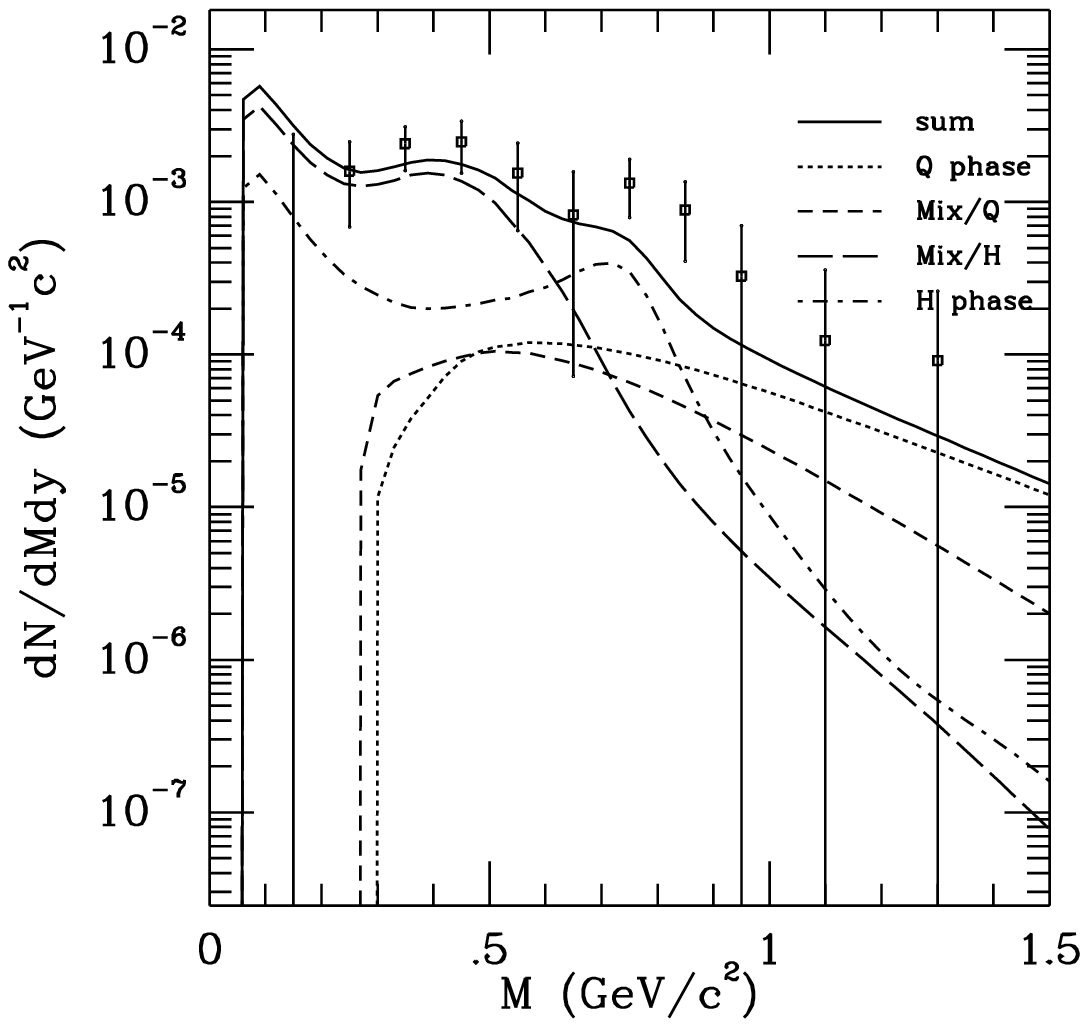}
\end{minipage}
%%\end{center}
\caption{\label{Mspectrum}
(a) Possible scenarios of chiral restoration in $m_\rho - m_{a_1}$
plane, (b)  the dilepton spectra corresponding to them (CERES data are
shown with background from hadronic decays subtracted); (c) the most 
favorable case D is shown in details, with contribution from different
stages. The dominant one is clearly the long-dashed one, or hadronic
part of the mixed phase.}
\end{figure}

  A point I want to make though
is that ``dropping  $m_\rho(T)$''  are not consistent with chiral
  symmetry restoration, unless  {\it modifications of its chiral partner}
 $a_1$ follow. Relation between the two were made
  e.g. in the contents of Weinberg-type sum rules
     \cite{KS_94}. Both states should become
identical at $T_c$, so   $m_\rho(T_c)=
  m_{A_1}(T_c)$. 
  Important role of $a_1$ for production of photons and dileptons was 
   discussed in  \cite{XSB,SX_where,Song_93,SKG_94}: we have used rather general
expressions recently derived in \cite{SYZ_96} \footnote{
In order to explain why $a_1$ is important, let us go ``backward in time'':
it is the first hadronic resonance which
may be excited in a collision of a photon, real or virtual, with a pion.
}
Possible scenarios of how chiral restoration may proceed are  shown in
the $m_\rho -   m_{A_1}$ plane in  Fig.\ref{Mspectrum}(a).
For example, path G corresponds to ref.\cite{Pisarski_95}
, while C corresponds to Brown-Rho scaling. The
corresponding
dilepton spectra are
 shown in  Fig.\ref{Mspectrum}(b):
  the variant D, with 
     $m_\rho(T_c)= m_{A_1}(T_c)\sim (1/2) m_\rho(0)$, does the best
     job, and Fig.\ref{Mspectrum}(c) we show contribution of separate 
stages in this scenario. The $A_1$ contribution is mainly at small  $M_{e+e-}$,
where  CERES acceptance is low and background rather high.
Still, with some effort one may be able to dig out 
the $A_1$ contribution, using other information (especially the $p_t$
distribution). Note finally, that none of the scenarios mentioned
above happen to violate  current upper limit
on the direct photons are set by WA80 \cite{WA80_seattle}.

\begin{figure}[h]
\begin{center}
\includegraphics[width=8.cm]{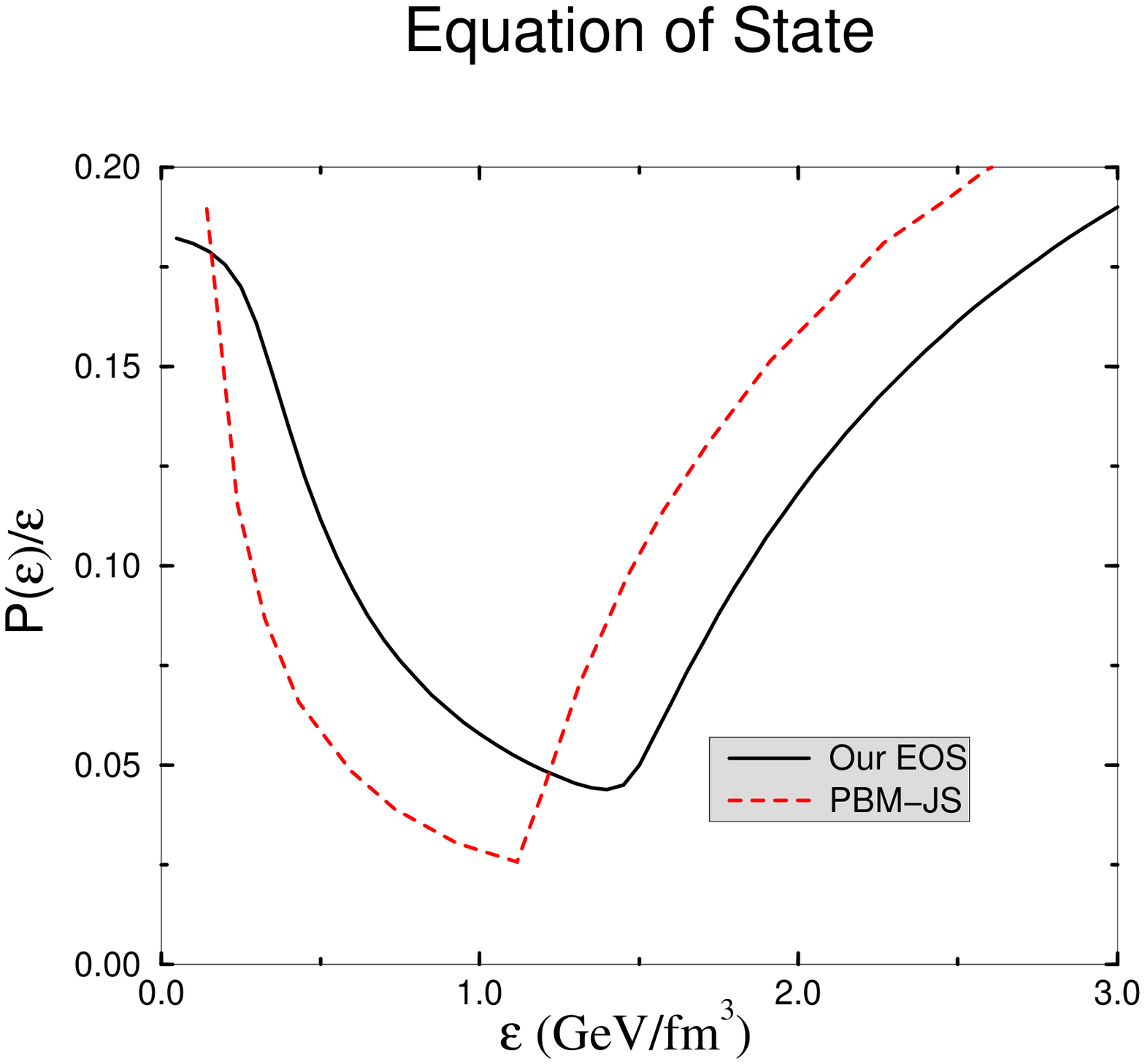}
\hspace{1cm}
\includegraphics[width=6.5cm]{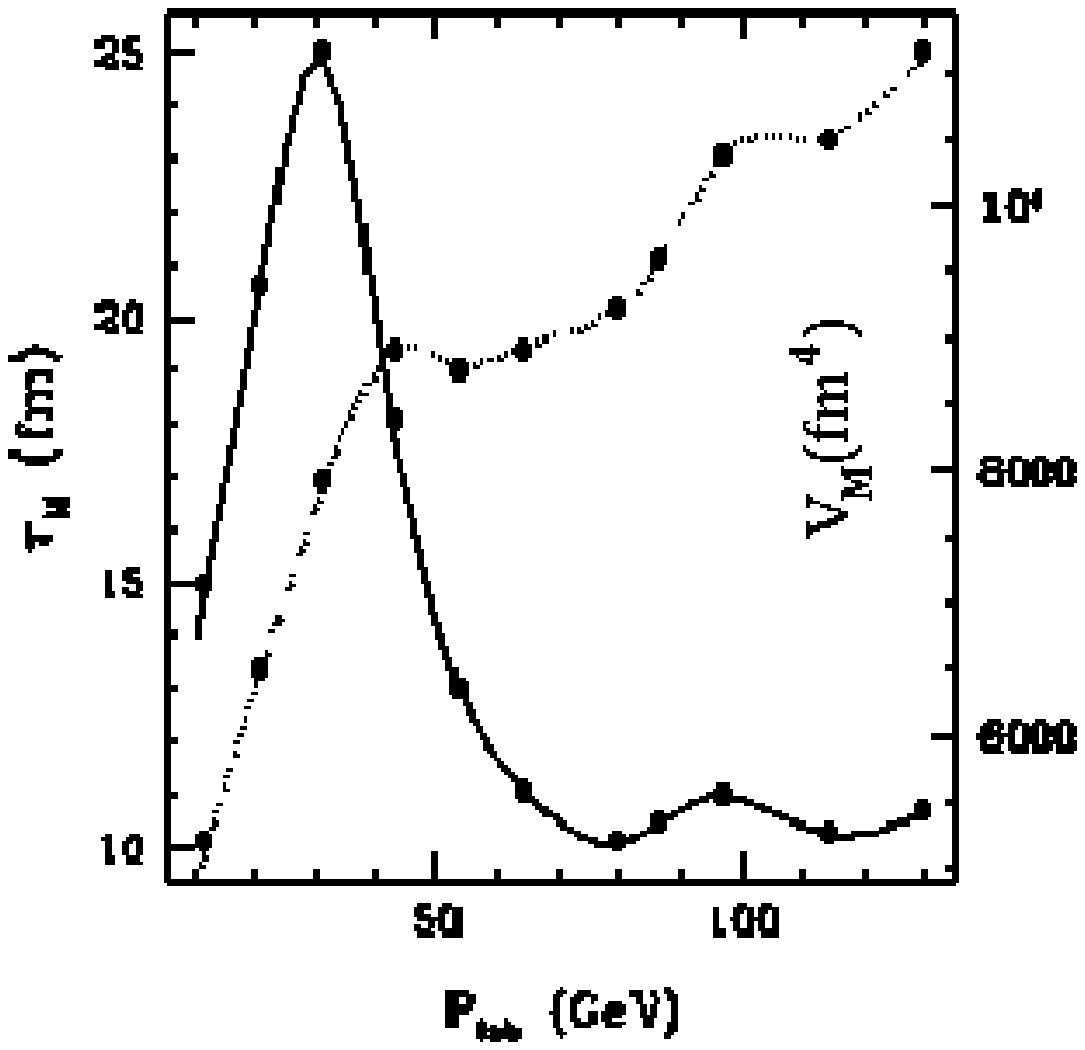}
\end{center}
\caption{\label{eos}
(a) The EOS in hydro-relevant coordinates for resonance gas and QGP.
The solid line is for zero baryon number, the dashed line is for
$\mu_b=0.54 GeV$. The  minimum is the
``softest point'' discussed in the text.
(b) Solid line (and left scale) show the lifetime in the center, the
dashed line (and the right scale) shows the space-time volume occupied
by the mixed phase ($T\approx T_c$).
}\end{figure}

\begin{figure}[h]
\begin{center}
\includegraphics[width=7.cm,angle =-90]{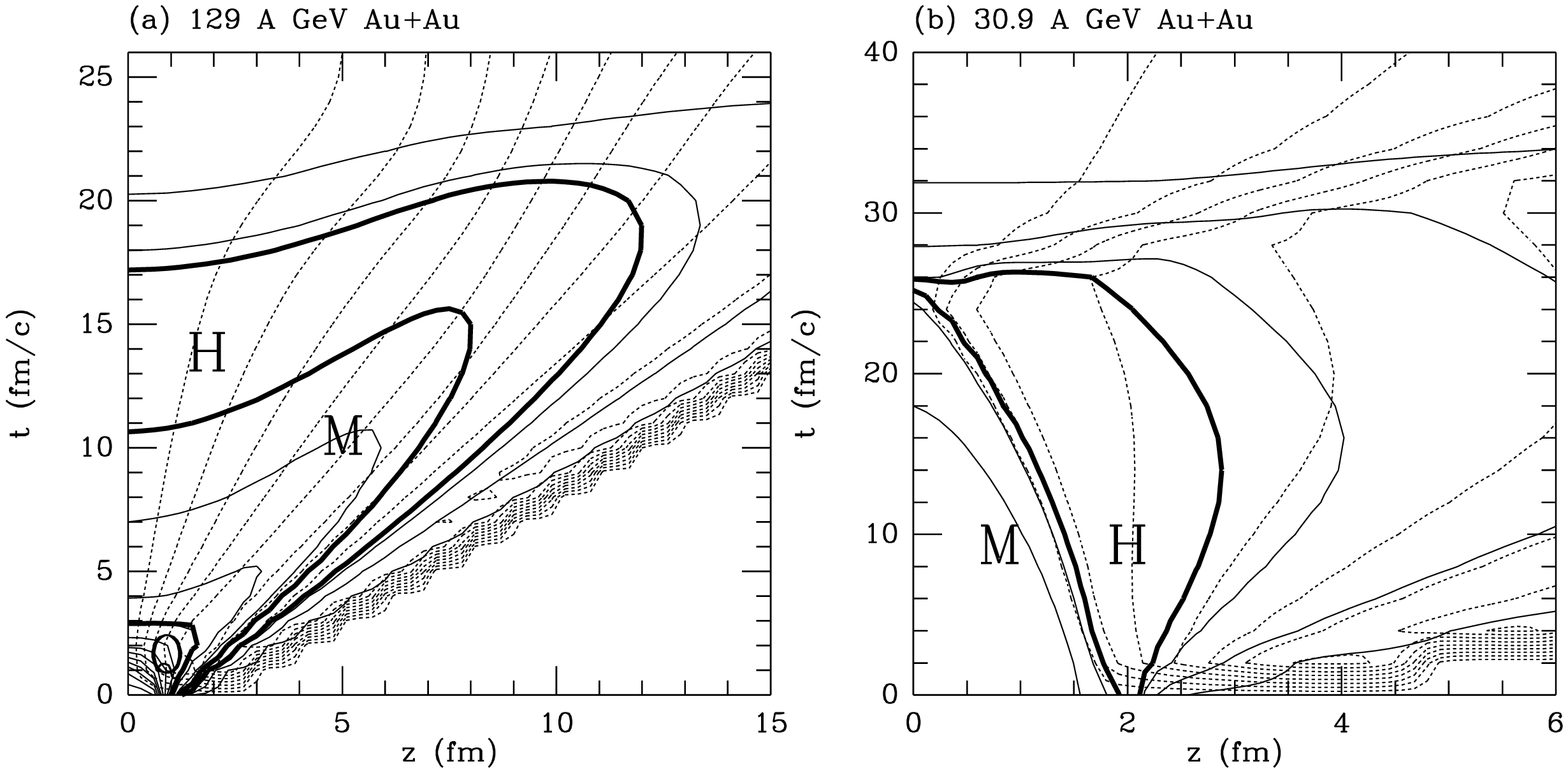}
%\hspace{1cm}
%\includegraphics[width=3.cm]{file=cases.eps}
\end{center}
\caption{\label{hydro}
Space-time picture of AuAu collisions at SPS energies (a) and at the
``softest point'' (b). Solid lines correspond to fixed energy density
while
the short-dash lines are contours of fixed longitudinal
velocity. Q,M,H stands for Quark, Mixed and Hadronic phases. Note 
qualitative difference between the two figures, as well as different 
time scales.
}\end{figure}

\section{ Flow, the ``softness'' of the EOS, and the fireball lifetime} 

  Our last topic deals with a more  straightforward
approach to observable signal of the QCD phase transition:  instead of
hunting for  ``dropping masses'' we look instead 
at the effect of
``dropping   pressure''.
The ratio $p/\epsilon$, pressure  normalized to the energy
density, is shown
in Fig.\ref{eos}(a) we
show how this ratio depends on $\epsilon$. The
existence of  the ``softest point'',   the
minimum of this ratio, is clearly seen. Our second point: in those
(hydro relevant) plot the curves with and without baryon density look
similar in many models: compare the two curves in Fig.\ref{eos}.

Effect of the
 ``softness'' of EOS at $T\approx T_c$ on transverse $radial$ flow 
 was discussed for a long time
\cite{SZ,transhydro}\footnote{The issue was strongly oversimplified:
in fact, transverse velocity is the integral over
the whole history, and initial ``softness" is only part of the story.
 Our hydro studies
have shown, that  the $mean$
flow velocity is not sensitive to EOS at early stages, although
there are modifications of the shape of its $ditribution$. } .
More radical (and more controversial)
 idea was proposed in \cite{HS_95}, where it was suggested that
 it can affect 
the $longitudinal$ expansion and thus the global lifetime of the
excited system in some energy window. In this workshop we had a
detailed talk by D.Rischke, where he has shown that
the effect of ``softness''
of EOS increases the lifetime at RHIC energies by about factor 2.

  It was shown in \cite{HS_95}  that, due to softness of the EOS,
 there exists a $window$ 
of collision energies in which the  
secondary
acceleration of matter is completely $impossible$.
Thus, $if$ stopping does occur, a very long-lived fireball should be formed. 

  To study  this idea semi-quantitatively, a simple
1-fluid relativistic hydrodynamic model was used\footnote{
The major uncertainties  are in the initial 
conditions: we assumed that $half$ of the total energy 
is stopped.}
Our lattice-based EOS shown (in unusual form) 
in Fig.\ref{eos}(a) have smooth crossover
in a $narrow$  temperature interval $  \Delta T \sim 5$ MeV
at $ T_c =160$ MeV the energy density jumps by about
an order of magnitude.

As the collision energy is scanned down, 
from 200 A GeV (SPS) to 
10 A GeV (AGS), we found three radically different scenarios:
{\bf(I)} At SPS one starts in the QGP phase,   therefore 
 longitudinal
explosion quickly restore  
ultra-relativistic longitudinal 
motion, even if stopping a la Landau takes place, and
 expansion resembles the  Bjorken picture,
% in spite of the assumed stopping
(see
  Fig.\ref{hydro} (a)): 
{\bf(II)}  As the initial energy density  passes the
softest point, the QGP phase disappears and so does rapid longitudinal
expansion.  Instead we find a {\it slow-burning}  fireball.
For the heavy nuclei and initial conditions we 
discuss, the burning front moves mostly in the
compressed longitudinal direction (see
  Fig.\ref{hydro} (b)).
 The total lifetime of the fireball appears to be nearly as long as
 predicted for RHIC, but of course it is kind of upper bound on the effect.
{\bf(III)} At still lower energies burning process becomes more spherical, and
ordinary expansion is developed. The main result, shown in
  Fig.\ref{eos}(b), is  a significant $peak$ in a lifetime.   

Remarkably enough, these completely different scenarios
lead to not-so-different spectra! With little fitting of the initial
conditions, they can be made 
consistent with available data. Thus, one-particle spectra alone is
not enough, and details about the space-time picture has to come from
analysis of flow and HBT.

How one can test these unusual predictions
 and try to locate the ``softest point''
 experimentally?
First of all, it seems  that it is above the AGS domain\footnote{ Preliminary
 results from the E802 AGS experiment
  reported  studies of HBT  which indicate significant 
 growth of lifetime for the most central Au Au collisions \cite{E802}.
},  so one
should rather
 scan downward at SPS. The following ideas
 are
discussed in literature:
(i) Look for   maximal lifetime
 (or minimum of  the HBT
   parameter $\lambda$) \cite{HS_95}; (ii)
 Look for  the {\it minimum} of  the ``directed'' flow 
in the collision plane \cite{RG_96};
(iii) Look for the nearly isotropic
distribution of  dileptons, produced in the long-lived fireball \cite{HS_95}.

\section*{Acknowledgments}
This talk is based on works done
my collaborators,
 especially with T.Schaefer and C.M.Hung. I benefited from
multiple discussion of the topics involved with G.Brown.
During the workshop strong interaction with
fellow co-organizers N.Christ,
M.Creutz, M.Gyulassy
and S.Kahana has shaped the interesting discussions we had.
This work is partially supported by the 
US Department
 of Energy under Grant No. DE-FG02-88ER40388.

%\bibliographystyle{unsrt}
%\bibliographystyle{plain}
%\bibliography{rev,matter}

\end{document}